\documentclass{article}
\usepackage{amsmath}
\usepackage{graphicx}
\usepackage{amssymb}
\usepackage{subfigure}
\usepackage[margin=1in]{geometry}

\newcommand{\beq}{\begin{equation}}
\newcommand{\eeq}{\end{equation}}
\newcommand{\bea}{\begin{eqnarray}}
\newcommand{\eea}{\end{eqnarray}}

\def\cp{{\rm CP}}

\newcommand{\gsim}{\lower.7ex\hbox{$\;\stackrel{\textstyle>}{\sim}\;$}}
\newcommand{\lsim}{\lower.7ex\hbox{$\;\stackrel{\textstyle<}{\sim}\;$}}

\newcommand{\bi}{\begin{itemize}}
\newcommand{\ei}{\end{itemize}}

\usepackage{color}
\usepackage{hyperref} 
\hypersetup{linktocpage=true}
\usepackage[all]{hypcap}
\hypersetup{
   bookmarks=true,         % show bookmarks bar?
   unicode=false,          % non-Latin characters in AcrobatÍs bookmarks
   pdftoolbar=true,        % show AcrobatÍs toolbar?
   pdfmenubar=true,        % show AcrobatÍs menu?
   pdffitwindow=false,     % window fit to page when opened
   pdfstartview={FitH},    % fits the width of the page to the window
   pdftitle={My title},    % title
   pdfauthor={Author},     % author
   pdfsubject={Subject},   % subject of the document
   pdfcreator={Creator},   % creator of the document
   pdfproducer={Producer}, % producer of the document
   pdfkeywords={keyword1} {key2} {key3}, % list of keywords
   pdfnewwindow=true,      % links in new window
   colorlinks=true,       % false: boxed links; true: colored links
   linkcolor=blue,          % color of internal links
   citecolor=magenta,        % color of links to bibliography
   filecolor=magenta,      % color of file links
   urlcolor=cyan           % color of external links
}

\usepackage{fancyhdr}
\begin{document}
\thispagestyle{empty}
\vspace*{-22mm}
\begin{flushright}
UND-HEP-10-BIG\hspace*{.08em}06\\
TUM-HEP-803/11
%MPP-2009-??\\
%Release Candidate 3\\
%\today
\end{flushright}
\vspace*{10mm}
\hypersetup{backref=true,bookmarks}

\vspace*{10mm}

\begin{center}
{\Large {\bf\boldmath 
Conclusions from CDF Results on  CP Violation 

\vspace*{3mm}

in $D^0 \to \pi^+\pi^-, \, K^+K^-$ and Future Tasks}}
\vspace*{10mm}

{\bf Ikaros\,I.\ Bigi$^a$, Ayan\ Paul$^a$, Stefan\ Recksiegel$^b$}\\
\vspace{4mm}
{\small
$^a$ {\sl Department of Physics, University of Notre Dame du Lac}\\
%\vspace*{-.8mm}\\
{\sl Notre Dame, IN 46556, USA}\vspace{1mm}
%{\sl email: ibigi@nd.edu} \\

$^b$ {\sl Physik Department, Technische Universit\"at M\"unchen,
D-85748 Garching, Germany}}

\vspace*{10mm}

{\bf Abstract}\vspace*{-1.5mm}\\
\end{center}
Within the Standard Model (SM) one predicts both {\em direct} and {\em indirect} CP violation in 
$D^0 \to \pi^+\pi^-$, $K^+K^-$ transitions, although the effects are tiny: 
Indirect CP asymmetry cannot exceed ${\cal O}(10^{-4})$, probably even ${\cal O}(10^{-5})$;  
direct effects are estimated at not larger than $10^{-4}$. 
At $B$ factories direct and indirect 
asymmetries have been studied with $\langle t \rangle /\tau_{D^0} \simeq 1$; no CP asymmetry 
was found with an upper bound of about 1\%.  CDF has shown 
intriguing data on CP violation in $D^0 \to \pi^+\pi^-$ [$K^+K^-$] with 
$\langle t \rangle /\tau_{D^0} \simeq 2.4$ [$2.65$]. 
Also, CDF has not seen any CP violation. For {\em direct} CP asymmetry, CDF has
a sensitivity similar to the combination 
of the $B$ factories, yet for {\em indirect} CP violation it yields a significantly smaller sensitivity of 
$a_{\cp}^{\rm ind}=(-0.01 \pm 0.06_{stat} \pm 0.05_{syst})$\% due to it being based on longer decay times.  
New Physics models (NP) like Little Higgs Models with 
T-Parity (LHT) can produce an {\em indirect} CP asymmetry up to 1\%; CDF's findings thus cover the upper range of 
realistic NP predictions $\sim 0.1 - 1$\%. One hopes that LHCb and a Super-Flavour Factory will probe the lower 
range down to $\sim 0.01$\%. Such {\em non}-ad-hoc NP like LHT cannot enhance {\em direct} CP violation significantly over the SM level 
in $D^0 \to \pi^+\pi^-$, $K^+K^-$ and $D^{\pm} \to \pi^{\pm}K^+K^-$
transitions, but others might well do so. 

\noindent

\newpage

\tableofcontents

%%%%%%%%%%%%%%%%%
\section{Introduction}
%%%%%%%%%%%%%%%

The existence of hadrons with charm as a strong stable quantum number was predicted by some courageous 
theorists to keep the analogy between quark and lepton families, reproduce the observed suppression of strangeness changing neutral currents and even save the renormalization of the Standard Model (SM). They also fell in the 
expected mass range and showed the predicted preference for decaying to strange hadrons with non-strange hadrons 
being suppressed by $\sim$ sin$\theta_C$. At that time most of the community saw little hope of finding 
manifestations of New Physics (NP) in charm transitions. Only a few physicists have advocated the search of NP in 
charm decays despite these successes of the SM, mainly because SM weak phenomenology is so 
`dull' with very slow $D^0 - \bar D^0$ oscillation, little CP violation and with rare decays dominated by Long Distance 
(LD) dynamics. 

Some possible hints of New Physics (NP) have appeared in charm physics. Compelling evidence 
for $D^0-\bar D^0$ oscillations has been presented by Belle, BaBar and CDF \cite{D0obs}. The HFAG has 
combined the results allowing for CP violation \cite{HFAGCHARM}\footnote{Up to date results can be found in the \href{http://www.slac.stanford.edu/xorg/hfag/charm/CHARM10/results_mix+cpv.html}{HFAG} website}: 
\bea
x_D = \frac{\Delta M_D}{\Gamma_D} = \left( 0.63 ^{+0.19}_{-0.20}\right) \% \; &,& \; 
y_D = \frac{\Delta \Gamma _D}{2\Gamma_D} = \left( 0.75 \pm 0.12\right) \% \\
\left |\frac{q}{p}  \right | = 0.91 ^{+0.18}_{-0.16} \; &,& \;  
\phi   = \left( - 10.2 ^{+9.4}_{-8.9}\right) ^o
\label{DOSCDATA}
\eea 
Oscillations happen if $x_D \neq 0$ and/or $y_D \neq 0$; indirect CP violation needs oscillations with  
$|q/p| \neq 1$ and/or $\phi \neq 0$. 

The observation of $D^0-\bar D^0$ oscillations seems established, while the relative size of $x_D$ and 
 $y_D$ is not clear yet. {\em Before} these experimental results in 2007, most authors had argued that the 
 SM predicts $x_D$, $y_D$ $\leq 3\times 10^{-4}$ --- yet not all. In 1998, 
 $x_D$, $y_D$ $\leq 10^{-2}$ was listed, admittedly as a {\em conservative} SM bound \cite{VAR98}, 
 together with a question: How can one rule out that the SM can not produce 
 $10^{-6} \leq r_D \leq 10^{-4}$ (corresponding to $x_D$, $y_D$ $\sim 10^{-3} - 10^{-2}$). In 2000 and 2003, an SM prediction obtained from an operator product expansion (OPE) yielded $x_D$, $y_D$ $\sim {\cal O}(10^{-3})$ 
 \cite{DUAL};  later a more sophisticated OPE analysis was done 
with similar results \cite{LENZ}. Alternatively 
 in 2001 and 2004, an SM prediction on $D^0-\bar D^0$ oscillations was based on $SU(3)$ breaking mostly in the phase space for $y_D$ and then from a dispersion relation for $x_D$ \cite{FALK}.
 
While the present experimental results on $x_D$ and $y_D$ can be accommodated within some available theoretical 
SM estimates --- and no non-zero CP asymmetry has been seen yet --- the observation of $D^0 - \bar D^0$ oscillations has `wetted' 
the appetite on thinking of NP in charm decays. The authors of Refs. \cite{BLUM,DMB,DKdual} 
suggest that NP could make a noticeable impact on charm decays.  In particular in \cite{DKdual} such has been 
analyzed, if future data confirm that $x_D$ indeed falls in the range of $0.5$\% and $1$\%. While SM 
long distance dynamics could accommodate a value in that range, it could not be ruled out that NP could contribute 
half or quarter value of $x_D$ considering the reasonable theoretical uncertainty in the predictions from 
long distance dynamics. Yet indirect CP violation in $D^0 - \bar D^0$ oscillations would provide us a clear signature 
for the manifestation of NP dynamics.

In neutral $K$ and $B_d$ decays CP violation has been observed for more than 45 years: since 1964 and 1999 
indirect and direct asymmetries have been established in $K_L$ transitions, and since 2001 (and later) in $B_d$ decays. They are (at least mostly) in agreement with SM flavour dynamics. No such asymmetries have been 
found in up-type --- $u$, $c$ and $t$ --- quark transitions. CP asymmetries cannot be as large in $D$ decays as in 
$B$ ones, the former cannot be even close to the latter. 
Yet observable CP violations are less suppressed for SM dynamics for $charm$ than for $up$ or $top$ quarks; the same 
conclusion is likely to hold for $charm$ forces in NP scenarios.  
 
In the next section we sketch SM CP phenomenology for charm and in Sect.\ref{CDF} explain the conclusions 
from CDF's data on CP violation in $D^0 \to \pi^+\pi^-/K^+K^-$ about restricting the parameter space 
of viable models of NP in 
a {\em non}-trivial way; in Sect.\ref{NPCPV}  
we discuss signals about a class of NP in CP violation, mostly of Little Higgs Models with T-parity, and future tasks   before our outlook in Sect.\ref{SUMM}. 

%%%%%%%%%%%%%%%%%
\boldmath
\section{CP Violation for $D$ Decays in SM Dynamics}
\unboldmath
\label{SMCPV}
%%%%%%%%%%%%%%%

SM flavour dynamics (with six quarks) create direct and indirect CP violation in $D$ transitions, but only 
with a small magnitude. Furthermore the theoretical predictions suffer from sizable uncertainties as discussed below.

There are three portals through which CP violation can enter in observable ways in two- and three-body final 
states:  
\begin{enumerate}
\item 
$|q| \neq |p|$; in `wrong-charged' {\em semi}-leptonic decays of neutral $D$ mesons,  
$D^0 - \bar D^0$ oscillations are the only source in the SM. One finds a time {\em in}dependent asymmetry 
\beq
A_{SL}^{\cp}(D^0 \to l^-X) = 
\frac{\left| \frac{p}{q}\right|^2 - \left| \frac{q}{p}\right|^2}{\left| \frac{p}{q}\right|^2 + \left| \frac{q}{p}\right|^2} \; .  
\eeq
The rate into wrong-charged leptons is time-dependent and tiny since it is given by $(x_D^2 + y_D^2)/2$ irrespective of what dynamics generate $x_D$ and $y_D$. Within the SM this manifestation of indirect CP violation is also tiny: 
\beq
A_{SL}^{\cp}(D^0 \to l^-X)|_{\rm SM} =  {\cal O}(10^{-3}) \; .  
\eeq
\item 
Direct CP asymmetries in non-leptonic decays can surface in $\Gamma (D\to f) \neq \Gamma (\bar D \to \bar f)$ 
in a time-independent way for different final states. 
\item 
`Tertium datur': there is a third portal in non-leptonic $D^0$ decays where CP
  violation can occur as given by 
$${\rm Im}\left( \frac{q}{p}\frac{T(\bar D^0 \to f)}{T(D^0 \to f)}\right)\cdot {\sin}\Delta M_D t$$where for simplicity $f$ is assumed to be a CP eigenstate. CP violation affects $q/p$ in $\Delta C=2$ dynamics and 
$T(\bar D^0 \to f)/T(D^0 \to f)$ in $\Delta C=1$ forces; the latter will create a difference between different 
final states like for $\pi^+\pi^-$ vs. $K^+K^-$. 

\end{enumerate}

As discussed below these three classes are not zero in the SM, but tiny and suffer large theoretical and 
experimental uncertainties in SM predictions.

%%%%%%%%%%%%%%
\subsection{SM Indirect CP Violation in Non-leptonic Charm Decays }
\label{SMINCPV}
%%%%%%%%%%

Indirect CP effects for $D^0$ in {\em non}-leptonic decays can have time-dependence 
due to oscillations parametrized by $x_D = \Delta M_D/\Gamma _{D^0}$ and 
$y_D = \Delta \Gamma_{D^0}/2\Gamma _{D^0}$; the indirect CP violating parameter can be approximated by implementing $x_D$, $y_D$ $\ll 1$ applied to the leading $D^0 \to K_S\phi$, where direct CP violation is very 
unlikely: 
\bea
\nonumber a^{\rm ind}_{NL}(D^0 \to K_S\phi; t) \equiv \frac{\Gamma (D^0 \to K_S\phi; t)-\Gamma (\bar D^0 \to K_S\phi; t)}{\Gamma (D^0 \to K_S\phi; t)+ \Gamma (\bar D^0 \to K_S\phi; t)}\simeq 
\label{AINDCPV1} \\
\simeq \frac{t}{\tau _{D^0}}\left[ y_D\left(\left| \frac{q}{p} \right| 
-  \left| \frac{p}{q} \right|        \right)\cos 2 \phi   -  
x_D\left(\left| \frac{q}{p} \right| 
+  \left| \frac{p}{q} \right|        \right)\sin 2 \phi      \right] \; ; 
\label{AINDCPV2}
\eea 
$y_D$ and $x_D$ could be as `large' as $0.01$ -- but also close to $0.005$ or less, 
in particular for one of the two. The theoretical SM estimates for $x_D$ and $y_D$ are hardly more than guesses and the theoretical uncertainties there are even larger than the experimental ones, since the SM predictions depend strongly on our 
theoretical treatment of LD dynamics for $x_D$ and $y_D$ (or its lack thereof) and our treatment 
of the extraction of the CKM phase as it enters charm decays \cite{DUAL,DKdual}:
\beq 
\sin 2 \phi \sim 10^{-3}
\eeq
\beq
a^{\rm ind}_{NL}(t)|_{\rm SM} \leq  {\rm several}\cdot 10^{-5}
\cdot \frac{t}{\tau _{D^0}}  \; , 
\eeq
where the oscillation's strength expressed through $y_D$ and $x_D$ --- or its weakness --- has been incorporated in this number. It  
depends sensitively on non-perturbative parameters \cite{DUAL} that are somewhat larger than 
one would assume if not suggested by experimental values for $x_D$ and $y_D$. For later discussion we mention that at the same time 
NP could naturally create a value for $x_D \simeq 0.01$, yet $y_D$ would be unlikely to be sensitive to NP. Using 
the experimental values of $|q_D/p_D|$ and $\phi$ stated in Eq.(\ref{DOSCDATA}) one gets 
\beq
|a^{\rm ind}_{NL}(t)|_{\rm exp}| \leq 1\% \; . 
\eeq

%%%%%%%%%%%%%%
\subsection{On Time Evolution with Indirect CP Asymmetries}
\label{TIMEEVOL}
%%%%%%%%%%

As expressed in Eq.(\ref{AINDCPV1}) one calculates the difference over the sum of partial rates. However 
one measures the BR$(D^0 \to f)$ vs.\ BR$(\bar D^0 \to f)$; i.e., $\Gamma (D^0 \to f)/\Gamma _{D^0}$ vs.  
$\Gamma (\bar D^0 \to f)/\Gamma _{\bar D^0}$. If oscillation happens to have $\Delta M_D \neq 0$, yet 
$\Delta \Gamma_D =0$, there is only one time scale, namely given by $\Gamma _{D^0}= \Gamma _{\bar D^0}$. However in the general case one has $\Delta \Gamma _D \neq 0$. Therefore the two mass eigenstates have 
two different lifetimes, and the two flavour eigenstates $D^0$ and $\bar D^0$ represent different combinations 
of $D_1$ and $D_2$: 
\bea
\Gamma (D^0(t) \to f) &\propto& \frac{1}{2}\cdot |T(D^0 \to f)|^2 \cdot G_f(t) \; , 
\nonumber \\ 
G_f(t) &=& Ae^{-\Gamma_1t} + Be^{-\Gamma_2t} + (C{\rm cos}\Delta M t + D{\rm sin}\Delta M t )e^{-\bar  \Gamma t}\\
\Gamma (\bar D^0(t) \to f) &\propto& \frac{1}{2}\cdot |T(\bar D^0 \to f)|^2 \cdot \bar G_f(t) \; , 
\nonumber \\
\bar G_f(t) &=& \bar Ae^{-\Gamma_1t} + \bar Be^{-\Gamma_2t} + 
(\bar C{\rm cos}\Delta M t + \bar D{\rm sin}\Delta M t )e^{-\bar \Gamma t} 
\eea
with $\bar \Gamma = (\Gamma_1 + \Gamma_2)/2$. The general expressions are given by
\bea
A&=& \frac{1}{2}\left( 1+ \left|\frac{q}{p}\frac{T(\bar D^0 \to f)}{T(D^0 \to f)}  \right|^2 \right)+
{\rm Re}\left( \frac{q}{p}\frac{T(\bar D^0 \to f)}{T(D^0 \to f)}  \right) \\
B &=& \frac{1}{2}\left( 1+ \left|\frac{q}{p}\frac{T(\bar D^0 \to f)}{T(D^0 \to f)}  \right|^2 \right)-
{\rm Re}\left( \frac{q}{p}\frac{T(\bar D^0 \to f)}{T(D^0 \to f)}  \right) \\
C &=& 1- \left|\frac{q}{p}\frac{T(\bar D^0 \to f)}{T(D^0 \to f)}  \right|^2 \; , \; 
D = -2{\rm Im}\left( \frac{q}{p}\frac{T(\bar D^0 \to f)}{T(D^0 \to f)}  \right)
\eea
\bea
\bar A&=& \frac{1}{2}\left( 1+ \left|\frac{p}{q}\frac{T(D^0 \to f)}{T(\bar D^0 \to f)}  \right|^2 \right)+
{\rm Re}\left( \frac{p}{q}\frac{T(D^0 \to f)}{T(\bar D^0 \to f)}  \right) \\
\bar B &=& \frac{1}{2}\left( 1+ \left|\frac{p}{q}\frac{T(D^0 \to f)}{T(\bar D^0 \to f)}  \right|^2 \right)-
{\rm Re}\left( \frac{p}{q}\frac{T(D^0 \to f)}{T(\bar D^0 \to f)}  \right) \\
\bar C &=& 1- \left|\frac{p}{q}\frac{T(D^0 \to f)}{T(\bar D^0 \to f)}  \right|^2 \; , \; 
\bar D = -2{\rm Im}\left( \frac{p}{q}\frac{T(D^0 \to f)}{T(\bar D^0 \to f)}  \right)
\eea
Without direct CP violation in the form $|T(D^0 \to f)| = |T(\bar D^0 \to f)|$ 
 --- like for $D \to K_S\phi$ --- one has 
\bea
\frac{q}{p}\bar \rho (f) = \left| \frac{q}{p}\right| e^{2i\tilde \phi } \; &,& \; 
\frac{p}{q}\frac{1}{\bar \rho (f)} = \left| \frac{p}{q}\right| e^{-2i\tilde \phi} \\
A= \frac{1}{2}\left( 1+ \left|\frac{q}{p} \right|^2 \right)+ \left| \frac{q}{p} \right| {\rm cos}2\tilde \phi 
   \; &,& \; 
\bar A= \frac{1}{2}\left( 1+ \left|\frac{p}{q}  \right|^2 \right)+
\left| \frac{p}{q}\right| {\rm cos}2\tilde \phi  \\
B= \frac{1}{2}\left( 1+ \left|\frac{q}{p} \right|^2 \right)-
 \left| \frac{q}{p} \right| {\rm cos}2\tilde \phi    \; &,& \; 
\bar B= \frac{1}{2}\left( 1+ \left|\frac{p}{q}  \right|^2 \right)-
\left| \frac{p}{q}\right| {\rm cos}2\tilde \phi \\
C = 1- \left|\frac{q}{p} \right|^2 \; &,& \; \bar C = 1- \left|\frac{p}{q} \right|^2 \\
D=-2\left| \frac{q}{p} \right| {\rm sin}2\tilde \phi    \; &,& \; \bar D = +2\left| \frac{p}{q}\right| {\rm sin}2\tilde \phi 
\eea
with $\tilde \phi = \phi - \alpha_f$, where $\alpha_f$ describes CP violation in $\Delta C =1$.  Therefore 
\bea
G_f(t) =&& e^{-\Gamma_1t}\left[ \frac{1}{2}\left( 1+ \left|\frac{q}{p} \right|^2 \right)+ 
\left| \frac{q}{p} \right| {\rm cos}2\tilde \phi\right] + e^{-\Gamma_2t}\left[  \frac{1}{2}\left( 1+ \left|\frac{q}{p} \right|^2 \right)-
 \left| \frac{q}{p} \right| {\rm cos}2\tilde \phi \right]   
\nonumber \\
&&+ e^{-\bar  \Gamma t}
\left[ \left( 1- \left|\frac{q}{p} \right|^2\right){\rm cos}\Delta M t 
-2\left| \frac{q}{p} \right| {\rm sin}2\tilde \phi \; {\rm sin}\Delta M t \right] \\
\bar G_f(t) =&& e^{-\Gamma_1t}\left[ \frac{1}{2}\left( 1+ \left|\frac{p}{q} \right|^2 \right)+ 
\left| \frac{p}{q} \right| {\rm cos}2\tilde \phi\right] + e^{-\Gamma_2t}\left[  \frac{1}{2}\left( 1+ \left|\frac{p}{q} \right|^2 \right)-
 \left| \frac{p}{q} \right| {\rm cos}2\tilde \phi \right]   
\nonumber \\
&&+ e^{-\bar  \Gamma t}
\left[ \left( 1- \left|\frac{p}{q} \right|^2\right){\rm cos}\Delta M t 
+2\left| \frac{p}{q} \right| {\rm sin}2\tilde \phi \; {\rm sin}\Delta M t \right] \; ;
\eea
i.e., the time evolutions of rates for $D^0 \to f$ and $\bar D^0 \to f$ are controlled by three time scales: two by the 
widths $\Gamma_1$ and $\Gamma_2$ of the two mass eigenstates $D_1$ and $D_2$ and the third one by 
$(\Gamma_1+ \Gamma_2)/2$ from the interference of $D^0 \to f$ and $\bar D^0 \to f$. Their weights depend 
on CP violation from $|q/p|\neq 1$ and $\tilde \phi \neq 0, \pi$. 

To calibrate the rates for $D^0/\bar D^0 \to f$ one can compare it for a decay where a CP asymmetry is 
very unlikely even with NP, namely $D^0 \to K^-\pi^+$ vs. $\bar D^0 \to K^+\pi^-$: 
\bea
\Gamma (D^0(t) \to f_{\rm no\, CPV}) \propto \frac{1}{4}K^2
\left[ e^{-\Gamma_1t} + e^{-\Gamma_2t} + 2e^{-\bar \Gamma t}  \right]
\\
\Gamma (\bar D^0(t) \to \bar f_{\rm no\, CPV}) \propto \frac{1}{4}K^2
\left[ e^{-\Gamma_1t} + e^{-\Gamma_2t} + 2e^{-\bar \Gamma t}  \right]
\eea
with $K=|T(D^0(t) \to f_{\rm no\, CPV})|= |T(\bar D^0(t) \to \bar f_{\rm no\, CPV})|$. Therefore 
\bea
\Gamma_{\rm calib}(D^0 \to f)) = \frac{\Gamma(D^0 \to f)}{\Gamma(D^0 \to f_{\rm no\, CPV})} = 
\frac{2G_f(t)}{K^2\left[ e^{-\Gamma_1t} + e^{-\Gamma_2t} + 2e^{-\bar \Gamma t}   \right]}
\\
\Gamma_{\rm calib}(\bar D^0 \to f) = \frac{\Gamma(\bar D^0 \to f)}{\Gamma(\bar D^0 \to \bar f_{\rm no\, CPV})} =
\frac{2\bar G_f(t)}{K^2\left[ e^{-\Gamma_1t} + e^{-\Gamma_2t} + 2e^{-\bar \Gamma t}   \right]}
\eea
With this we find through first order in $y_D$, $x_D$: 
\beq
R=\frac{\Gamma_{\rm calib}(D^0 \to f)-\Gamma_{\rm calib}(\bar D^0 \to f) }
{\Gamma_{\rm calib}(D^0 \to f)+ \Gamma_{\rm calib}(\bar D^0 \to f) } = 
\frac{e^{\bar \Gamma t}\left[ G_f(t)- \bar G_f(t)\right] }{e^{\bar \Gamma t}\left[ G_f(t)+\bar G_f(t)\right] }
+{\cal O}(y_D^2)
\eeq
with 
\bea
e^{\bar \Gamma t} G_f(t) &=& 2\bar \Gamma t  \left( 
1 -y_D\left|\frac{q}{p} \right|{\rm cos}2\tilde \phi 
-x_D \left| \frac{q}{p} \right| {\rm sin}2\tilde \phi \right) \\
e^{\bar \Gamma t} \bar G_f(t) &=& 2 \bar \Gamma t \left( 
1 -y_D \left|\frac{p}{q} \right|{\rm cos}2\tilde \phi 
+x_D \left| \frac{p}{q} \right| {\rm sin}2\tilde \phi  \right) \; ; 
\eea
thus 
\beq
R \simeq -\bar \Gamma t \left[ y_D {\rm cos}2\tilde \phi \left( \left| \frac{q}{p} \right|   -\left| \frac{p}{q} \right| \right)
+ x_D {\rm sin}2\tilde \phi \left( \left| \frac{q}{p} \right|  + \left| \frac{p}{q} \right| \right)
\right]\,.
\eeq
If $D^0 - \bar D^0$ oscillations show $y_D$ to be measurable, one has two different lifetimes 
of the two mass eigenstates $D_1$ and $D_2$. Yet for $y_D \ll 1$, the ansatz used by CDF applies for indirect CP asymmetries in non-leptonic transitions holds to first order in $y_D$:  
\beq
A_{\rm indir. \, CPV} \simeq \frac{\langle t \rangle }{\bar \tau} a_{\rm indir. \, CPV}= 
\langle t \rangle \frac{\left( \Gamma_1 + \Gamma_2 \right)}{2}a_{\rm indir. \, CPV}
\eeq
The same conclusions hold when direct CP violation can surface in  
$D^0 \to \pi^+\pi^-$, $K^+K^-$.

%%%%%%%%%%%%%%
\subsection{Direct CP Violation}
\label{SMDIRCPV}
%%%%%%%%%%

Purely direct CP asymmetries need two 
amplitudes $A_{1,2}$ with non-zero weak phase $\Delta \phi$ and strong phases $\Delta \alpha$; thus they exhibit direct time-independent asymmetries: 
\beq
a^{\rm dir}_{D \to f} = \frac{|A_f|^2 - |\bar A_f|^2}{|A_f|^2 + |\bar A_f|^2} = 
\frac{-2|A_1||A_2|\sin \Delta \alpha \times \sin \Delta \phi}
{|A_1|^2 + |A_2|^2 + 2|A_1||A_2|\cos \Delta \alpha \times \cos \Delta \phi}
\eeq  
The strong phases $\Delta \alpha$ mostly come from LD dynamics as do a large part of the two 
$|A_{1,2}|$, of which we do not have good quantitative theoretical
control. There is a large uncertainty already within the SM. 

Nevertheless Cabibbo-suppressed non-leptonic decays of charm mesons $D \to f$ contain the needed ingredients, namely two amplitudes with different weak and in general strong phases. Fortunately for $D^0 \to \pi^+\pi^-$ 
decays they can be obtained from other measurements, namely comparing branching ratios of $D^0 \to \pi^+\pi^-$, $\pi^0\pi^0$ and 
$D^{\pm} \to \pi^{\pm}\pi^0$. 

Such direct CP asymmetries have been quoted in the literature as SM predictions for Cabibbo suppressed 
non-leptonic $D$ decays of about $10^{-4}$ or less for the final state $\pi^+\pi^-$: 
\beq
a^{\rm dir}_{D^0 \to \pi^+\pi^- ;{\rm SM}} \leq 10^{-4} \; . 
\eeq
Our current calculations confirm this and we can reproduce numbers very close to the upper limit mentioned above.  For direct charm asymmetries it can occur in the form of 
Penguin diagrams like for $B$ transitions. For $B$ decays they are well defined, since the leading contribution --- like 
for $b\to tW \to d$ --- is given by a {\em local} operator because $m_t \gg m_b$. Yet for charm decays the leading diagram for $c \to s W \to u$ looks like a Penguin graph, but does not represent a local operator, since $m_s < m_c$. 
Therefore one has to allow for even larger uncertainties than usual. Therefore we want to address 
explicitly the uncertainties in the SM predictions.

%%%%%%%%%%%%%%
\boldmath
\subsubsection{Revisiting Direct CP Asymmetries in $D \to hh$ within the SM}
\unboldmath
\label{SMDPP}
%%%%%%%%%%%%%%

SM contributions to direct CP asymmetries in $D \to hh$ come primarily from the tree level current-current  and gluonic penguins operators. The latter, even though loop suppressed, get enhanced in the matrix elements due to a factor of $\sim m_{\pi}^2/m_u^2$ for $D^0\to \pi^+\pi^-$; it is smaller in $D^0 \to K^+ K^-$. 
The ansatz applied to $B \to \pi K, \pi\pi$ decays \cite{BBNS} has been extended to Cabibbo-suppressed $D$ decays 
in Ref.\cite{GKN}. These calculations of amplitudes include current-current, the gluonic penguin and the electric dipole operators.  In our preparation to include LHT contributions in $D \to hh$, we had to take this calculation a step further. The reasons for our labour will be evident when we discuss LHT contributions to $a_\cp^{dir}$ in 
Sect.\ref{DIRCPIND}. The details of this calculation shall not be discussed in this paper, but will be elucidated on in a future work. We shall only list the salient features of this calculation here. One has to keep in mind that 
non-perturbative effects are much less under theoretical control in $D \to hh$ than for $B\to hh$. 
\begin{itemize}
\item We include twelve $\Delta C=1$ operators pertinent to the process: two current-current, four gluon penguins,  four electroweak penguins, an electric dipole and a chromomagnetic dipole operator. 
\item All calculations are done in next-to-leading order in $\alpha_s$.
\item A renormalization scale of $\mu=1$ GeV was chosen.
\item  Only a limited set of hadronic non-perturbative parameters have been included which are necessary for inclusion of all the operators mentioned above.
\item Both hard gluon exchange contributions involving spectator quarks and weak annihilation contribution have been ignored. 
\end{itemize}
The SM predictions of $D \to hh$ decay rates reach within about a factor of two of the experimental values; 
this is about as much as one can expect considering the theoretical uncertainties.

%%%%%%%%%%%%
\boldmath
\section{CDF's CP Searches in $D^0$ Decays}
\unboldmath
\label{CDF}
%%%%%%%%% 

The two factories at Belle and BaBar have established $D^0 - \bar D^0$ oscillations and have searched data for  
direct and indirect CP asymmetries; no positive signals have been found for 
$\langle t\rangle \simeq \tau _{D^0}$ in $D^0 \to \pi^+\pi^-$, $K^+K^-$ averaged by Belle/BaBar: 
\bea
\langle A^{\rm BF}_{\cp}(D^0 \to \pi^+\pi^-)\rangle &=& a^{\rm dir}_{\cp} + a^{\rm ind}_{\cp} = (+0.11 \pm 0.39) \%  \\ 
\langle A^{\rm BF}_{\cp}(D^0 \to K^+ K^-)\rangle &=& a^{\rm dir}_{\cp} + a^{\rm ind}_{\cp} \simeq (- 0.24 \pm 0.24) \% \; .
\label{BFRESULT}
\eea 
CDF has now shown data on CP asymmetry in $D^0 \to \pi^+\pi^-$, $K^+K^-$ \cite{CDFRESULT}: 
\bea
\langle A^{\rm CDF}_{\cp}(D^0 \to \pi^+\pi^-)\rangle &=&  
(+0.22 \pm 0.24_{stat} \pm 0.11_{syst}) \%    
\\
\langle A^{\rm CDF}_{\cp}(D^0 \to K^+ K^-)\rangle &=& (- 0.24 \pm 0.22_{stat} \pm 0.10_{syst}) \%  \; .
\label{CDFRESULT}
\eea
CDF's data has exhibited an overall similar sensitivity as the average of the two $B$ factories. It shows that neither 
Belle/BaBar nor CDF data have the sensitivity to get even close to the level of the SM predictions 
around $10^{-4}$ or less for direct and indirect CP violation. 

One  important feature of CDF's data is their acceptance for $D^0$ decays at significantly larger times than at the 
$B$ factories, namely 
$\langle t\rangle = (2.40 \pm 0.03) \tau _{D^0}$ [$=(2.65 \pm 0.03) \tau _{D^0}$] 
for $D^0 \to \pi^+\pi^-$ [$K^+K^-$]; therefore 
\bea
\langle A^{\rm CDF}_{\cp}(D^0 \to \pi^+\pi^-)\rangle &\simeq& 
a^{\rm dir}_{\cp}( \pi^+\pi^-) + 2.40 \times a^{\rm ind}_{\cp} \\
\langle A^{\rm CDF}_{\cp}(D^0 \to K^+ K^-)\rangle &\simeq& 
a^{\rm dir}_{\cp}(K^+K^-) + 2.65 \times a^{\rm ind}_{\cp}\,.
\eea
As a moderate input from theory, as explained in Sect.\ref{NPCPV}, NP models
can generate a much larger 
impact on {\em indirect} CP violation, but hardly any for a {\em direct} CP asymmetry. Assuming 
$a^{\rm dir}_{\cp} =0$, CDF finds a much larger sensitivity for indirect CP violation than Belle/BaBar:
\bea
a^{\rm ind}_{\cp}(D^0 \to \pi^+\pi^-) |_{\rm CDF} &=& (+0.09 \pm 0.10_{stat} \pm 0.05_{syst}) \%
\\ 
a^{\rm ind}_{\cp}(D^0 \to K^+ K^-)  |_{\rm CDF} &=& (-0.09 \pm 0.08_{stat} \pm 0.04_{syst}) \%
\\
a^{\rm ind}_{\cp}(D^0 \to \pi^+\pi^- / K^+ K^-)  |_{\rm CDF} &=& (-0.01 \pm 0.06_{stat} \pm 0.05_{syst}) \% 
\label{CDFIND} 
\eea
or 
\beq
|a^{\rm ind}_{\cp}(D^0 \to \pi^+\pi^- / K^+ K^-)  |_{\rm CDF} < 0.14\% \; {\rm at} \;  95\% \, {\rm CL} \; . 
\eeq
As explained below, the experimental values of $x_D$, $y_D$, $|q/p|$ and $\phi_D$ tell us that 
$|a^{\rm ind}_{\cp}| \leq 1\%$ at best. Non-ad-hoc NP models can produce such values. Belle and BaBar 
bounds can barely enter that regime -- but CDF bounds can enter the upper part of the regime.

%%%%%%%%%%%%%%%%%%%%
\section{CP Violation due to New Physics}
\label{NPCPV}
%%%%%%%%%%%

Charm dynamics have been viewed as a signature success of the SM going back it its discovery. Yet few authors 
have seen charm decays as a chance to find NP due to the SM's dullness in its weak phenomenology.  
To say it differently: while contributions from NP will not be large or even sizable, they will not have to deal with a `background' from SM's phenomenology. 

NP's impacts on weak charm dynamics are often discussed in isolation of other flavours and in an  
ad-hoc model fashion. Instead we want to discuss scenarios that are primarily
motivated by the weak-electric phase 
transition; yet they can have implications for flavour dynamics. More specifically, we consider the
Littlest Higgs Models with T-Parity designed to deal with the `little hierarchy' problem: interestingly they do {\em not} represent  
`Minimal Flavour Violation' models in $K$, $B$ and $D$ decays; i.e., one can find non-SM predictions 
in rare decays and CP asymmetries. 

As mentioned before, LHT can have a large impact on $\Delta C =2$ dynamics which is sketched below. Therefore  we have next analyzed 
LHT's impact on $\Delta C =1$ dynamics in $D^0 \to \gamma \gamma/\mu^+\mu^-$ \cite{PBR1} and 
$D \to l^+l^-X$ \cite{PBR2}. The results are that LHT models hardly have any noticeable impact.  
Significant enhancement over SM contribution can be seen only in forward-backward (FB) asymmetries. However, large $O(10\%)$ asymmetry can only be seen in the CP asymmetry of the FB asymmetry while the pure FB asymmetry is limited to $O(1\%)$ \cite{PBR2}. 

In the analysis to follow on direct CP violation from NP models like LHT, we continue to use the same parameter set as has been used in the previous works;  details can be found in \cite{PBR1,PBR2}. This parameter set was also used in the earlier analysis of indirect CP violation and $D^0 - \bar{D}^0$ oscillation in \cite{BBBR}.

%%%%%%%%%%%%%%
\boldmath
\subsection{Indirect CP Violation on $D^0 \to h^+h^-$ from LHT Models}
\unboldmath
\label{INDIRLHT}
%%%%%%%%%%

As shown in Ref.\cite{BBBR}, for some regions in parameter space LHT can produce
\beq
|a^{\rm ind,LHT}_{\cp}| \leq 1 \%
\label{ONE}
\eeq
assuming that LHT generates most of the observed $D^0 - \bar D^0$ oscillations (otherwise less). 
In such a scenario one gets $|{\rm sin}\phi _{weak, NP}|$ up to unity and therefore a large CP asymmetry in 
wrong-charge 
leptons, larger by at least two orders of magnitudes than the SM. 

The combination of Belle/BaBar results just enters the regime of Eq.(\ref{ONE}). 
Yet CDF's data, Eq.(\ref{CDFIND}), establish a 
{\em very non-trivial} upper bound of about 
0.2\% from an interesting class of NP models motivated outside of flavour dynamics. 
LHT-like models \cite{PBR1} 
could generate a signal for indirect CP violation `just around the corner' for future measurements.

%%%%%%%%%%%%%%%%%%%%
\subsection{Direct CP Violation in LHT Models}
\label{DIRCPIND}
%%%%%%%%%

\begin{figure}[h!]
\includegraphics[width=16cm]{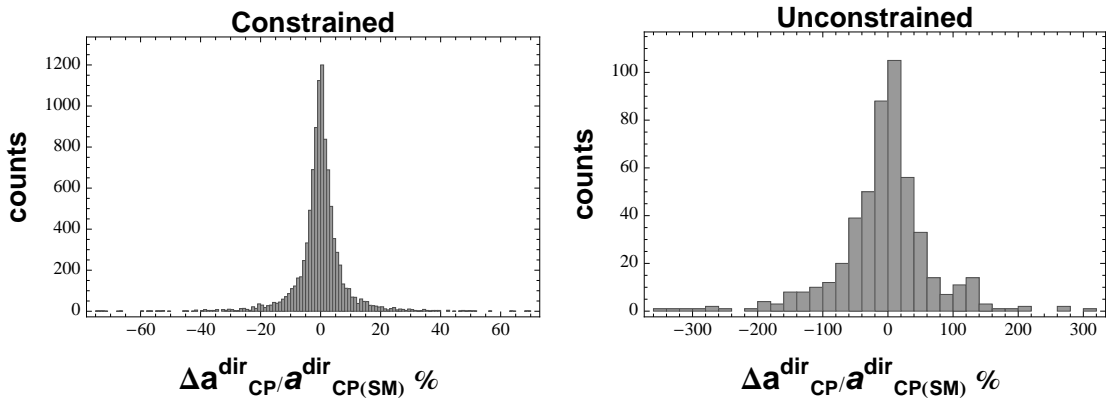}
\caption{Histogram of percentage increase in $a^{dir}_\cp$ due to LHT contributions over SM values for both the constrained and unconstrained set.}
\label{fig:ACP}
\end{figure}

Like in the SM, direct CP asymmetries in non-leptonic $D$ decays are generated from the interference of the 
SM tree level diagrams with one-loop diagrams from NP. In the LHT all the
one-loop $\Delta C=1$ diagrams are {\em local} operators as the internal states
are very heavy --- unlike for the SM. However, a closer analysis shows that LHT models cannot produce a larger 
enhancement to SM direct CP asymmetries unlike for the case of indirect CP violation; it will be evident from the following line of argument. Before plunging into the next few paragraphs, it is recommended for the enthusiastic reader to follow the scaling arguments that have been presented in \cite{PBR2} as that will be vital to our argument.

Firstly, enhancement to direct CP violation over the SM is expected from LHT-like models because of two reasons: (i) the presence of large phases in the new mixing matrices and (ii) presence of heavy mirror quarks which can produce effects akin to what has so commonly been seen due to the existence of top in the SM. However, this has to compete with the breaking scale suppression of $\frac{1}{64}(v^2/f^2)$ that the LHT loop functions will have relative to the SM loop functions, a factor of the $O(10^{-3})$.

In the SM the leading contribution is from the tree level current-current operator and the power enhanced gluon penguins. However, in the LHT the scenario is different. To begin with, due to T parity, there are no tree level contributions. The loop contributions can be divided into three types: 
\\

\begin{itemize}
\item gluon penguin operators (GPOs); 
\item electroweak penguin operators (EPOs); 
\item chromomagnetic and electric dipole penguin operators (CDPs). 
\end{itemize}
The GPOs scale as $\log(x)$. Here $x$ is the square of the ratio between the mass of the internal fermion to the mass of the internal gauge boson that exist as virtual states in the loop. In the SM $x\sim O(10^{-7} - 10^{-2})$. 
In the LHT $x\sim O(1)$. Hence from both SM and LHT GPOs receive the same contribution from loop functions. However, the LHT functions are suppressed by breaking scale as noted before. This results in very tiny contributions to the GPOs from LHT.

For the EPOs, the loop functions scale as $x$ or $x\log(x)$. Hence, EPOs receive much larger enhancement from LHT than GPOs. Despite the fact that EPOs come with a relatively smaller coupling of $\alpha$ compared to the 
$\alpha_s$ present in the GPOs, the huge enhancement due to the loop functions compensate for this. However at this point the unitarity of the mixing matrices also becomes important. Since the masses of the mirror quarks are not too hierarchical, the suppression from the unitarity of the mass mixing matrix is significant. It results in a comparable,  but not significantly larger enhancement to either the SM decay rate or the CP violating parameter. Of course, the CP violating parameter sees more enhancement than the decay rate as it probes the phases in the new mixing matrices which are large.

The CDPs also scale as $x$ but asymptotically reach a constant value for large $x$. They do show the possibility of getting large contributions from the LHT flavour sector. However, CDPs are both $\alpha_s/4\pi$ (or $\alpha/4\pi$) and colour suppressed. Hence contributions from them are not very significant.

All of this is quite contrary to what is seen in the SM and hence it is very easy to overlook the EPO contribution from NP. However, in any NP scenario which limits the presence of tree level processes, large enhancements of direct CP violation are not possible. We see at most an enhancement factor of two to $a_\cp^{dir}$ for the unconstrained set and 
${\cal O}(10\%)$ enhancement for the constrained set as seen in fig.\ref{fig:ACP} for $D^0\to\pi^+\pi^-$ and somewhat less for $D^0\to K^+K^-$.\footnote{cf \cite{PBR1,PBR2} for details of the parameter sets.}

One can understand with little work that NP cannot generate a large enhancement in direct CP asymmetries as it might happen for indirect CP violation. However some theorists have allowed sizable CP asymmetries Ð like one of authors of this paper. Yet a class of NP models Ð of which LHT models are a prominent example Ð shows no sizable enhancements. However, this Ôclose-to-zeroÕ result is due to several subtle features of NP models and the impact of QCD radiative corrections. One of the authors of this paper, in spite of being pleased with the knowledge gained, might despair that the absence of any sizable effect might make his work a waste of efforts on his side. Yet the other two authors (not having spent a lot of time on those tough calculations) consider even a zero result relevant for theory, if it implies subtle arguments like the scaling of NP contributions and the impact of QCD radiative corrections.

%%%%%%%%%%%%
\subsection{Future Tasks}
\label{FUT}
%%%%%%%%

Asymmetries from indirect CP violation have to be the same in $D^0 \to \pi^+\pi^-$ and $D^0 \to K^+K^-$ (and 
$D^0 \to K_S\phi$). If those asymmetries are different, they need a source for direct CP violation in 
$\Delta C=1$ processes --- and LHT-like models would become unlikely candidates for NP.

%%%%%%%%%%%%%%%%
\section{Summary}
\label{SUMM}
%%%%%%%%%%%

The existence of charm and the features of its dynamics have been seen ---
correctly so --- as a great success of 
our understanding of the SM. Nevertheless we should probe weak charm decays for manifestations of NP --- 
in particular in studies of CP violation in different areas. The SM generates small (although non-zero) asymmetries in 
wrong-charge semi-leptonic and in non-leptonic transitions, while NP models can produce much more sizable 
asymmetries than the SM. They are large relative to the SM effect, but not in absolute size; therefore the data have only now achieved the necessary sensitivities. 

The case of whether the SM can produce the observed values for $x_D$ (and
$y_D$) or whether NP is needed, has not been 
decided due to theoretical and experimental uncertainties. However the previous data on indirect CP violation in $D^0$ transitions had hardly entered the regime where {\em non-ad-hoc} NP models like LHT can generate observed effects. The CDF data implies a very non-trivial bound of 0.14 \% (at 95\% CL) for $|a^{\rm ind}_{\cp}(D^0 \to hh)|$. 
It also means that a manifestation of NP might be `around the corner' for future data.  

The pessimism about direct CP asymmetries derived from LHT-like NP models does not mean that searches for 
direct CP violation in $D$ decays are useless. Experimentalists should get the theorists' betting line, which can 
reduce the number of possible culprits --- but then probe CP invariance in $D\to 3h,\, 4h$.

%%%%%%%%%%%%
\section{Acknowledgements}
\label{ACK}
%%%%%%%%%%

We benifitted from our conversations with Luciano Ristori and Diego Tonelli from CDF. This work was supported by the NSF under the Grant No. PHY-0807959.

\fancyhf{}
 
\lhead{\uppercase{References}}
\cfoot{\thepage}

\end{document}